# Tighter and Stable Bounds for Marcum Q-Function

Jiangping Wang

*Abstract*—This paper proposes new bounds for Marcum Q-function, which prove extremely tight and outperform all the bounds previously proposed in the literature. What is more, the proposed bounds are good and stable both for large values and small values of the parameters of the Marcum Q-function, where the previously introduced bounds are bad and even useless under some conditions. The new bounds are derived by refined approximations for the $0^{th}$ order modified Bessel function in the integration region of the Marcum Q-function. They should be useful since they are always tight no matter the parameters are large or small.

## I. INTRODUCTION

Marcum Q-function was first used to deal with radar communications about almost half a century ago [1]. Since then, it has found many other applications, in particular the evaluation of error probability associated with communication systems [2, 3]. Currently Marcum Q-function is used to estimate various error probabilities. However, one popular expression of Marcum Q-function (1) is the generalized integration and its integrand involves $0^{th}$ order modified Bessel function, which render the difficulty for computation. For numerical computations and theoretical analyses, it is worth researching on the bounds of Marcum Q-function.

$$Q_1(a,b) = \int_b^\infty x\exp(-\frac{x^2+a^2}{2})I_0(ax)dx \tag{1}$$

Computing bounds of Marcum Q-function is always a challenging direction. In [3] the author proposed exponential-type bounds that have simple expression but are quite loose. In [4, 5] the alternative integral expressions are used to obtain new bounds which are both tighter than [1]. In [6] the author got the tighter bounds with simple expressions via a Geometric Approach. Up to now, the tightest bounds were proposed in [7], which fully utilized the characters of $0^{th}$ order modified Bessel function and inequality approach techniques. This paper mainly focuses on the tightness and robustness of bounds of Marcum Q-function. Thus, comparison to the bounds in [7] is a must. The bounds in [7] are extremely tight in most cases; however, it can be testified that the bounds in [7] are quite loose and even unbounded when arguments are getting smaller. Also the bounds get loose when arguments are large enough. This paper overcomes all these weaknesses and proposes new bounds in different cases by finding refined approximations for the $0^{th}$ order modified Bessel function.

Jiangping Wang is with the Electronic Engineering Department, Tsinghua University, Beijing, 100084, P. R. China (86-10-51534524; e-mail: wjp04ster@gmail.com).

In the case b>a, the integrand is monotonic decreasing in the region x>b apparently; thus we can find the upper and lower bounds of the integrand to approximate the integral (1). In case b<a, the integrand is firstly increasing and is then decreasing in the region x>b, but instead we can evaluate 1-Q(a,b), which is monotonic increasing in [0, b] and thus make the problem easier to solve.

In section two, the theoretical analysis is provided to derive the new bounds. Then the comparisons between the new bounds and previously proposed bounds in the literature are made in section three.

## II. MARCUM Q-FUNCTION BOUNDS

### A. Case $b \geqslant a$

Note that the bounds are in the interval [0, 1] that is in the case of the upper bounds one should consider min{1, upper bound}while in the case of the lower bounds one should consider max{0, lower bound}. It is common for computing bounds in the literature.

*1) Upper bounds:*

As for the function f(x) below, it is easy to prove that f(x) is monotonic decreasing.

$$f(x) = \frac{I_0(x)}{e^x + 3} \tag{2}$$

Since

$$I_0^{'}(x) = I_1(x)$$

$$\frac{df(x)}{dx} = \frac{I_1(x)(e^x+3) - I_0(x)e^x}{(e^x+3)^2}$$

$$= \frac{e^x[I_1(x)-I_0(x)] + 3I_1(x)}{(e^x+3)^2}$$

we can consider the function below.

$$g(x) = e^x[I_1(x) - I_0(x)] + 3I_1(x) \tag{3}$$

Obviously, g(x) is negative in the region x>0 from Fig. 1. So I can conclude that the f(x) (2) is monotonic decreasing in x>0, obtaining

$$\frac{I_0(x)}{e^x+3} \leq \frac{I_0(b)}{e^b+3}, \quad \forall x \geq b \tag{4}$$

Thus I get the following inequality

$$I_0(x) \leq \frac{I_0(b)(e^x+3)}{e^b+3}, \quad \forall x \geq b \tag{5}$$

It is easy to prove that

$$\frac{e^x}{e^b} > \frac{e^x+e^{-x}}{e^b+e^{-b}} > \frac{e^x+1}{e^b+1} > \frac{e^x+m}{e^b+m}, \quad \forall m>1, x>b \tag{6}$$

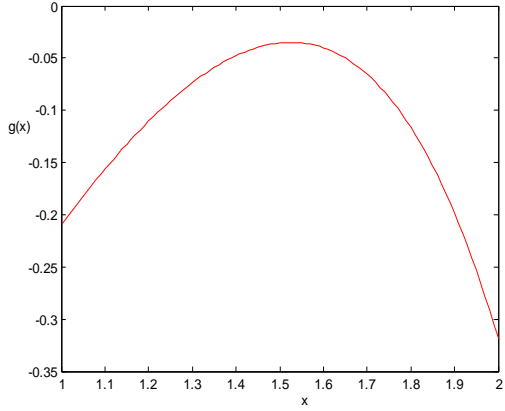

Fig. 1. The value of function g(x) in [1,2]

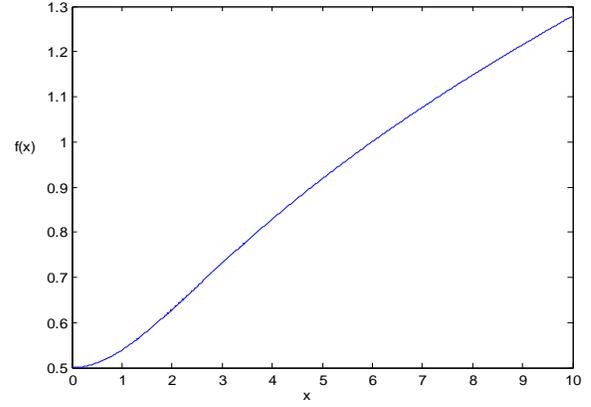

Fig. 2. The value of f(x) in [0, 10]

From (6), I work out the following inequality

$$I_0(x) \le \frac{I_0(b)(e^x+3)}{e^b+3} < \frac{I_0(b)e^x}{e^b}, \quad \forall x \ge b \quad (7)$$

In [7] the author used the following approximation [7, eq.(6)]

$$I_0(x) \le e^x \frac{I_0(b)}{\exp(b)}, \quad \forall x \ge b$$

With regard to (7), a new upper bound can be derived and is tighter than the bound proposed in [7]. From (1) and (7), the new upper bound is as follows:

$$Q_1(a,b) = \int_b^\infty x\exp(-\frac{x^2+a^2}{2})I_0(ax)dx$$

$$\le \frac{I_0(ab)}{\exp(ab)+3}\int_b^\infty x\exp(-\frac{x^2+a^2}{2})(\exp(ax)+3)dx$$

$$= \frac{I_0(ab)}{\exp(ab)+3}\{\exp(-\frac{(b-a)^2}{2})+a\sqrt{\frac{\pi}{2}}erfc(\frac{b-a}{\sqrt{2}})+3\exp(-\frac{a^2+b^2}{2})\} \quad (8)$$

Denote the new upper bound as UB1JP. In table I, the expression of UB1JP is shown, together with typical upper bounds that previously proposed in literature. The bound indicated as UB1A is proposed in [7], the bound UB1B is proposed in [4], the bound UB1C is proposed in [5] and the bound UB1D is proposed in [6].

*2) Lower Bounds:*

The function f(x) below is monotonic increasing in x>0, which can be easily proved with numerical computation from Fig. 2.

$$f(x) = \frac{xI_0(x)}{e^x - e^{-x}}$$

In x>b, the following inequality is satisfied.

$$\frac{xI_0(x)}{e^x - e^{-x}} \ge \frac{bI_0(b)}{e^b - e^{-b}} \quad (9)$$

Thus I can obtain that

$$I_0(ax) \ge \frac{bI_0(ab)}{e^{ab} - e^{-ab}} \frac{e^{ax} - e^{-ax}}{x}, \quad \forall x \ge b \quad (10)$$

From (2) (10), a new lower bound can be proposed as follows:

TABLE I
UPPER BOUNDS IN CASE b≥a

| | |
|---|---|
| UB1JP | $\frac{I_0(ab)}{\exp(ab)+3}\{\exp(-\frac{(b-a)^2}{2})$ $+a\sqrt{\frac{\pi}{2}}erfc(\frac{b-a}{\sqrt{2}})+3\exp(-\frac{a^2+b^2}{2})\}$ |
| UB1A | $\frac{I_0(ab)}{\exp(ab)}\{\exp(-\frac{(b-a)^2}{2})+a\sqrt{\frac{\pi}{2}}erfc(\frac{b-a}{\sqrt{2}})\}$ |
| UB1B | $\frac{b}{b-a}\exp(-\frac{(b-a)^2}{2})$ |
| UB1C | $\exp(-\frac{a^2+b^2}{2})I_0(ab)+a\sqrt{\frac{\pi}{8}}erfc(\frac{b-a}{\sqrt{2}})$ |
| UB1D | $[1-\frac{\arctan(b/a)}{\pi}]\exp[-\frac{(b-a)^2}{2}]$ $+\frac{\arctan(b/a)}{\pi}\exp(-\frac{a^2+b^2}{2})$ |

$$Q_1(a,b) = \int_b^\infty x\exp(-\frac{x^2+a^2}{2})I_0(ax)dx$$

$$\ge \frac{bI_0(ab)}{e^{ab}-e^{-ab}}\int_b^\infty \exp(-\frac{x^2+a^2}{2})[\exp(ax)-\exp(-ax)]dx$$

$$= \sqrt{\frac{\pi}{2}}\frac{bI_0(ab)}{e^{ab}-e^{-ab}}[erfc(\frac{b-a}{\sqrt{2}})-erfc(\frac{b+a}{\sqrt{2}})] \quad (11)$$

Denote the new lower bound as LB1JP.

In [7], the author used the inequality below to approach the $0^{th}$ order modified Bessel function [7, eq.(8)]

$$I_0(x) \ge \frac{I_0(b)b}{\exp(b)}\frac{\exp(x)}{x}, \quad \forall x \ge b \quad (12)$$

Accordingly, we can obtain that

$$\frac{\exp(x)-\exp(-x)}{\exp(b)-\exp(-b)} \ge \frac{\exp(x)}{\exp(b)}, \quad \forall x \ge b \quad (13)$$

and

$$I_0(x) \ge \frac{bI_0(b)}{e^b-e^{-b}}\frac{e^x-e^{-x}}{x} \ge \frac{I_0(b)b}{\exp(b)}\frac{\exp(x)}{x}, \quad \forall x \ge b \quad (14)$$

### TABLE II
### LOWER BOUNDS IN CASE b≥a

| | |
|---|---|
| LB1JP | $\sqrt{\dfrac{\pi}{2}}\dfrac{bI_0(ab)}{e^{ab}-e^{-ab}}[erfc(\dfrac{b-a}{\sqrt{2}})-erfc(\dfrac{b+a}{\sqrt{2}})]$ |
| LB1A | $\sqrt{\dfrac{\pi}{2}}\dfrac{bI_0(ab)}{e^{ab}}erfc(\dfrac{b-a}{\sqrt{2}})$ |
| LB1B | $\dfrac{b}{b+a}\exp(-\dfrac{(b+a)^2}{2})$ |
| LB1C | $\exp(-\dfrac{a^2+b^2}{2})I_0(ab)$ |
| LB1D | $[1-\dfrac{\arctan(b/a)}{\pi}]\exp[-\dfrac{a^2+b^2}{2}]$ $+\dfrac{\arctan(b/a)}{\pi}\exp(-\dfrac{(a+b)^2}{2})$ |

### TABLE III
### UPPER BOUNDS IN CASE b≥a

| | |
|---|---|
| UB2JP | $1-\dfrac{I_0(ab)}{e^{ab}+3}\{4\exp(-\dfrac{a^2}{2})$ $-\exp(-\dfrac{(b-a)^2}{2})-3\exp(-\dfrac{a^2+b^2}{2})$ $+a\sqrt{\dfrac{\pi}{2}}[erfc(-\dfrac{a}{\sqrt{2}})-erfc(\dfrac{b-a}{\sqrt{2}})]\}$ |
| UB2A | $1-\dfrac{I_0(ab)}{e^{ab}}\{\exp(-\dfrac{a^2}{2})-\exp(-\dfrac{(b-a)^2}{2})$ $+a\sqrt{\dfrac{\pi}{2}}[erfc(-\dfrac{a}{\sqrt{2}})-erfc(\dfrac{b-a}{\sqrt{2}})]\}$ |
| UB2D | $1-\dfrac{\arctan(b/a)}{\pi}\{\exp[-\dfrac{(a^2-b^2)^2}{2(a^2+b^2)}]$ $-\exp(-\dfrac{a^2+b^2}{2})\}$ |

That indicates that the newly proposed lower bound is tighter than the bound proposed in [7]. In table II, the bound LB1A is proposed in [7], the bound LB1B is in [4], LB1C is in [5] and LB1D is in [6].

### B. Case b<a

As mentioned in the introduction, since the integrand in (1) is not monotonic, we can focus on $1-Q_1(a,b)$ which is monotonic increasing. The following equation is satisfied.

$$Q_1(a,b)=1-\int_0^b x\exp(-\dfrac{x^2+a^2}{2})I_0(ax)dx \quad (15)$$

*1) Upper Bounds:*

To compute upper bound, from equation (15) we can try to derive the lower bound of second term of (15). From (4), we can get the inequality below.

$$I_0(x)\geq \dfrac{I_0(b)}{\exp(b)+3}(\exp(x)+3), \quad x\in[0,b] \quad (16)$$

With (2) (16), a new upper bound is obtained:

$$Q_1(a,b)=1-\int_0^b x\exp(-\dfrac{x^2+a^2}{2})I_0(ax)dx$$
$$\leq 1-\dfrac{I_0(ab)}{e^{ab}+3}\int_0^b x\exp(-\dfrac{x^2+a^2}{2})[\exp(ax)+3]dx$$
$$=1-\dfrac{I_0(ab)}{e^{ab}+3}\{4\exp(-\dfrac{a^2}{2})-\exp(-\dfrac{(b-a)^2}{2})-3\exp(-\dfrac{a^2+b^2}{2})$$
$$+a\sqrt{\dfrac{\pi}{2}}[erfc(-\dfrac{a}{\sqrt{2}})-erfc(\dfrac{b-a}{\sqrt{2}})]\} \quad (17)$$

Denote this new upper bound as UB2JP.

In [7] the author used the approximation for $0^{th}$ order modified Bessel function as follows:

$$I_0(x)\geq \dfrac{I_0(b)b}{\exp(b)}\exp(x), \quad x\in[0,b] \quad (18)$$

It is easy to prove that

$$\dfrac{\exp(x)+3}{\exp(b)+3}\geq \dfrac{\exp(x)}{\exp(b)}, \quad x\in[0,b] \quad (19)$$

Thus, the following inequality is satisfied

$$I_0(x)\geq \dfrac{I_0(b)}{\exp(b)+3}(\exp(x)+3)\geq \dfrac{I_0(b)b}{\exp(b)}\exp(x), \quad x\in[0,b] \quad (20)$$

From (20), it demonstrates that the new bound UB2JP is tighter than the bound in [7].

In table III, the bound UB2A is proposed in [7], UB2D is in [6]. Note that there are no proposed upper bound in case b<a in [4, 5].

*2) Lower Bounds:*

Similarly, we can work out the upper bound of $0^{th}$ order modified Bessel function in [0, b] instead.

It is worth remarking that in case b<a, the author of [7] took power function as the approximation for $0^{th}$ order modified Bessel [7, eq. (13)] instead of the inequality below.

$$I_0(x)\leq \dfrac{bI_0(b)}{\exp(b)}\dfrac{\exp(x)}{x}, \quad x\in[0,b] \quad (21)$$

The inequality (21) can be obtained by the function below which is monotonic in [0, b]

$$f_1(x)=\dfrac{xI_0(x)}{\exp(x)}$$

The reason is that the right term of (21) is infinite when x=0. However, here I propose another function that can be used to better approximate $0^{th}$ order modified Bessel function in [0, b]. Consider function

$$f(x)=\dfrac{xI_0(x)}{e^x-e^{-x}}$$

This function is monotonic increasing in x>0. Thus I can obtain the following inequality

$$I_0(x)\leq \dfrac{bI_0(b)}{\exp(b)-\exp(-b)}\dfrac{\exp(x)-\exp(-x)}{x}, \quad x\in[0,b] \quad (22)$$

It is easy to prove that

$$\lim_{x\to 0}\dfrac{bI_0(b)}{\exp(b)-\exp(-b)}\dfrac{\exp(x)-\exp(-x)}{x}=\dfrac{2bI_0(b)}{\exp(b)-\exp(-b)}$$

(23)

To make comparison with the approximation function proposed in [7], I can operate as follows:

Rice P.D.F is below:

$$R(x) = x\exp(-\frac{x^2+a^2}{2})I_0(ax) \quad (24)$$

Inequality (22) can derive the following inequality

$$R(x) \leq \frac{bI_0(ab)}{\exp(ab)-\exp(-ab)}[\exp(-\frac{(x-a)^2}{2})-\exp(-\frac{(x+a)^2}{2})], \quad x \in [0,b] \quad (25)$$

The inequality [7, eq. (13)] is below

$$R(x) \leq x\exp(-\frac{x^2+a^2}{2})\exp(\frac{log(I_0(ab))}{b}x), \quad x \in [0,b] \quad (26)$$

It is clear that (25) is tighter than (26) in [0, b] from Fig.3. What is more, the advantage of inequality (25) will be more obvious with the increase of arguments. Accordingly, a tighter lower bound can be derived as follows:

$$Q_1(a,b) \geq 1 - \frac{bI_0(ab)}{\exp(ab)-\exp(-ab)}\int_0^b[\exp(-\frac{(x-a)^2}{2})-\exp(-\frac{(x+a)^2}{2})]dx$$

$$= 1 - \frac{\sqrt{2\pi}bI_0(ab)}{\exp(ab)-\exp(-ab)}[erf(\frac{a}{\sqrt{2}})-\frac{1}{2}erf(\frac{a-b}{\sqrt{2}})-\frac{1}{2}erf(\frac{a+b}{\sqrt{2}})] \quad (27)$$

Denote this new lower bound as LB2JP. In table Ⅳ, the bound LB2A is in [7], LB2B is in [4], LB2C is in [5] and LB2D is in [6].

Up to now, this paper has fully proposed all the upper bounds and lower bounds in case b>a and b<a. From the theoretical analysis, I can prove that the newly proposed bounds are tighter than the bounds in [7]. Since the bounds proposed in [7] prove tighter than the previously provided bounds, it indicates that the bounds in this paper are tighter than the all the bounds in the literature. Furthermore, based on the following simulation and comparison, the newly proposed bounds in this paper are quite stable even when the argument a and b are relatively small or large. Thus it makes the new bounds more meaningful both for theory analysis and practical application.

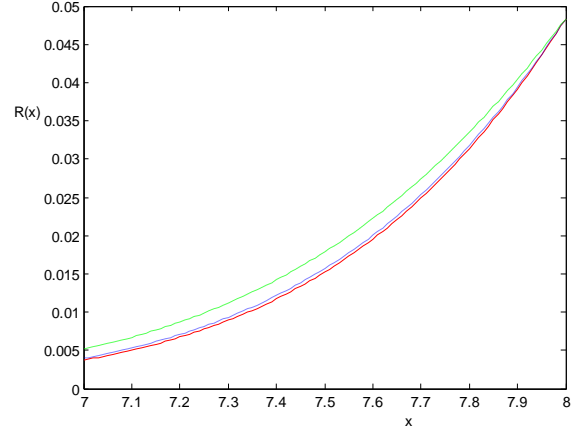

Fig. 3. The Rice probability distribution function value in [7, 8]. a=10, b=8. Red solid curve is the exact curve of Rice p.d.f. Blue dashed : the function in (25) proposed in this paper, Green dash dotted line: the function in (26) proposed in [7]

TABLE Ⅳ
LOWER BOUNDS IN CASE b≥a

| | |
|---|---|
| LB2JP | $1 - \frac{\sqrt{2\pi}bI_0(ab)}{\exp(ab)-\exp(-ab)}[erf(\frac{a}{\sqrt{2}}) - \frac{1}{2}erf(\frac{a-b}{\sqrt{2}}) - \frac{1}{2}erf(\frac{a+b}{\sqrt{2}})]$ |
| LB2A | $1 - \exp(-\frac{a^2-\varsigma^2}{2})\{\exp(-\frac{\varsigma^2}{2}) - \exp(-\frac{(b-\varsigma)^2}{2}) + \varsigma\sqrt{\frac{\pi}{2}}[erfc(-\frac{\varsigma}{\sqrt{2}}) - erfc(\frac{b-\varsigma}{\sqrt{2}})]\}$, $\varsigma \triangleq \frac{\log I_0(ab)}{b}$ |
| LB2B | $1 - \frac{a}{a-b}\exp(-\frac{(a-b)^2}{2})$ |
| LB2C | $\exp(-\frac{a^2+b^2}{2})I_0(ab)$ |
| LB2D | $1 - \frac{\arcsin(b/a)}{\pi}\{\exp[-\frac{(b-a)^2}{2}] - \exp[-\frac{(a+b)^2}{2}]\}$ |

## III.  COMPARISON

Here, I first compare the new bounds with the bounds proposed in [4, 5, 6] which are denoted as B, C and D for simple recognition. Then the special comparisons of the new bounds and the bounds in [7] are made by numerical computation with error tables.

The first comparisons are given in case a=1, 10 with respect to a>b and a≤b. In this way, the robustness of the bounds can be observed in respect to the increase of argument a.

The second comparison between new bounds and the bounds in [7] is special because the bounds in [7] are very tight overall. To make comparison, it will set relatively small and large arguments.

Denote the new bounds as JP and the bounds in [7] as A.

### A.  JP V.S. B&C&D

From Fig. 4-7, it is clear that the new bounds are much tighter than the bounds proposed in [4-6] in all kinds of cases. Furthermore, with the increase of argument a and b, the B bounds get so loose that they cannot estimate Marcum Q-function to some extent. The C bounds are even useless when a is a little larger. The D bounds are tight in the case b<a, but quite loose in the case b>a. So they are not stable enough. The bounds proposed in this paper denoted as JP are extremely tight in all cases and not getting loose with the increase or decrease of argument a and b.

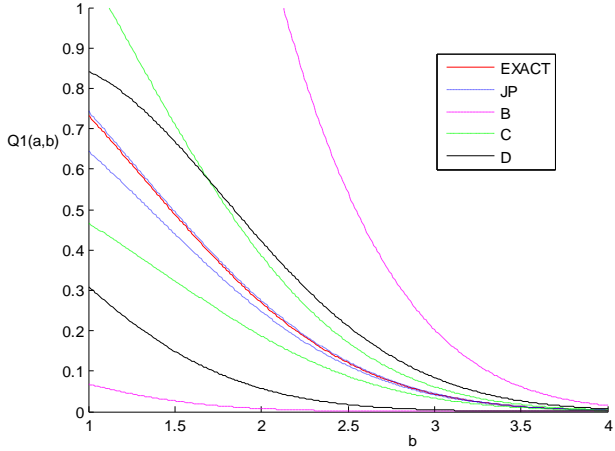

Fig. 4. Comparison between the function exact first order Marcum Q, JP(UB1JP&LB1JP), B, C, D when a=1 and b≥a.

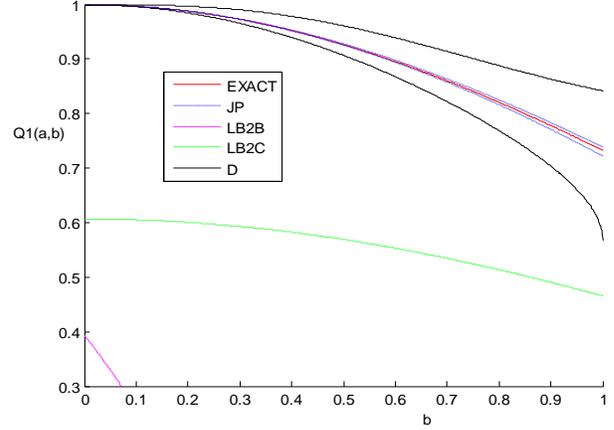

Fig. 6. Comparison between the function exact first order Marcum Q, JP(UB1JP&LB1JP), B, C, D where a=1 and b<a.

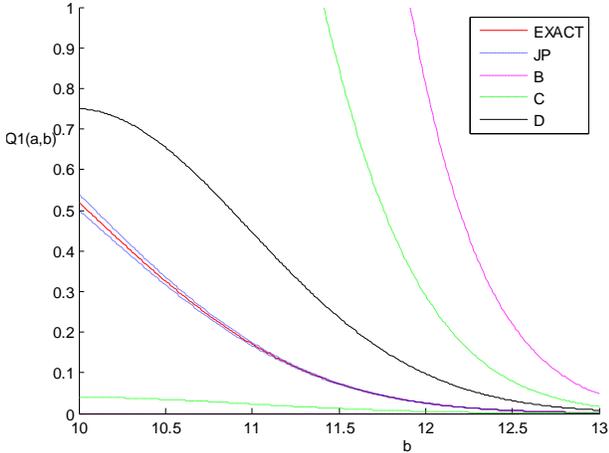

Fig. 5. Comparison between the function exact first order Marcum Q, JP(UB1JP&LB1JP), B, C, D where a=10 and b≥a.

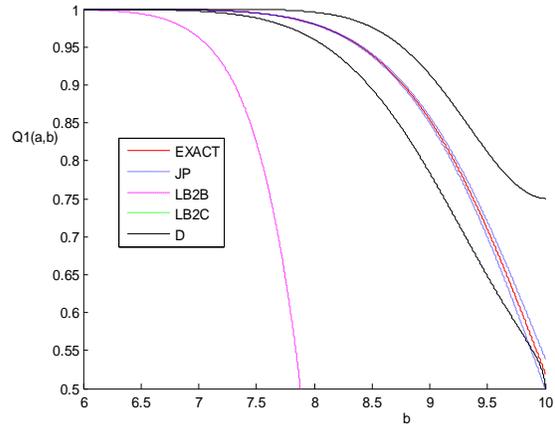

Fig. 7. Comparison between the function exact first order Marcum Q, JP(UB1JP&LB1JP), B, C, D where a=10 and b<a.

*B. JP V.S A*

This section focuses on the comparison of the new bounds (denoted as JP) and the tightest bounds (denoted as A) in the literature.

From Fig.8, when a is relatively small, the bounds proposed in [7] get loose even unbounded when b is close to a. In contrast, the new bounds in this paper are still quite tight in the same condition.

From Fig. 9, in the case b<a, the new lower bound is much tighter than LB2A proposed in [7].

From Fig. 10, in the case b<a, when a is relatively small, the new upper bound is much tighter than the bound in [7]. In table 5-8 present the error data of the new bounds UB1JP, LB1JP, UB2JP, LB2JP and the bounds in [7] UB1A，LB1A, UB2A, LB2A.

From Table5-8, it is clear that the new bounds in this paper outperform the bounds proposed in [7] and thus become the tightest bounds up to now. The stability is a significant factor that makes the new bounds quite robust and applicable.

IV. CONCLUSION AND DISCUSSION

Apart from tightness, another good character of the new bounds in this paper is robustness since they are not sensitive to the increase and decrease of argument a and b. The key techniques lie in the refined approach function for $0^{th}$ order modified Bessel function. All the bounds previously proposed seem not stable and tight enough and even become useless under some conditions. Though the bounds in [7] are quite tight in most cases, but they get unbounded when the argument is small, which make them not perfect enough. The highlight of this paper results from the stability of the new bounds. In some practical application where the robustness and tightness are significant, the new bounds in my paper may be quite useful.

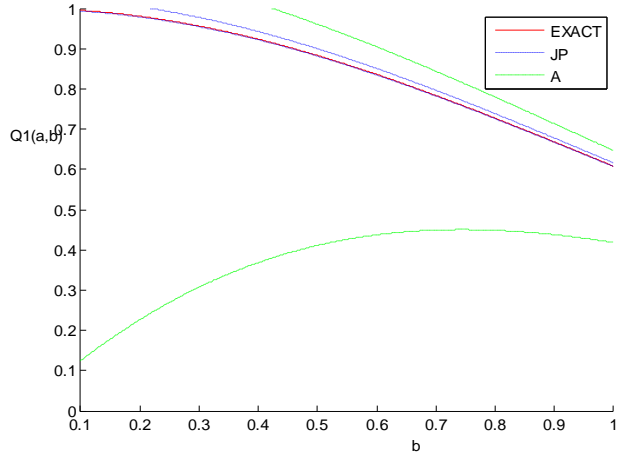

Fig. 8. Numerical results for Q1(a,b) and its bound JP and A versus b for the case of a=0.1 and b≥a.

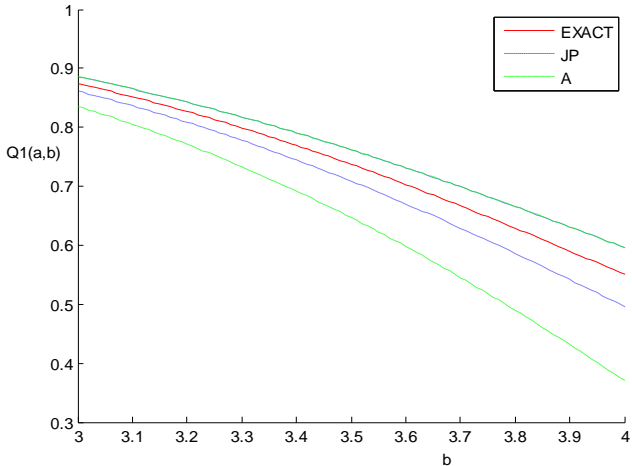

Fig. 9. Numerical results for Q1(a,b) and its bound JP and A versus b for the case of a=4 and b<a.

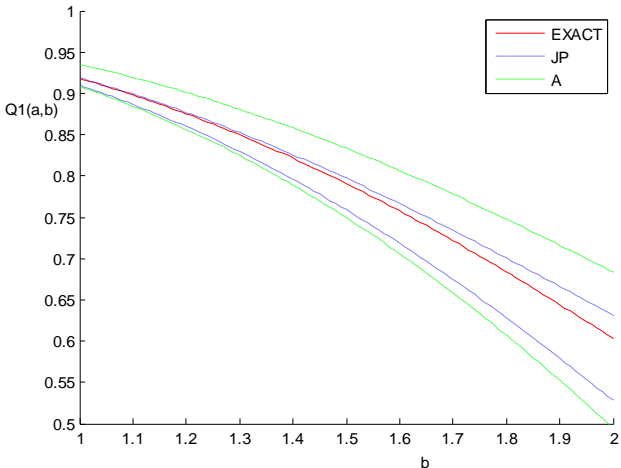

Fig. 10. Numerical results for Q1(a,b) and its bound JP and A versus b for the case of a=2 and b<a.

TABLE V
UPPER BOUND COMPARISON: a=0.1, b≥a

| b | Q1(a,b) | UB1JP | UB1JP (ε %) | UB1A | UB1A (ε %) |
|---|---|---|---|---|---|
| 0.1 | 0.99503 | 1.02133 | 2.6423 | 1.11416 | 11.9718 |
| 0.2 | 0.98029 | 1.00410 | 2.4284 | 1.08848 | 11.0360 |
| 0.3 | 0.95621 | 0.97762 | 2.2384 | 1.05381 | 10.2070 |
| 0.4 | 0.92348 | 0.94259 | 2.0689 | 1.01093 | 9.4697 |
| 0.5 | 0.88304 | 0.89997 | 1.9172 | 0.96085 | 8.8114 |
| 0.6 | 0.83602 | 0.85091 | 1.7810 | 0.90475 | 8.2212 |
| 0.7 | 0.78366 | 0.79665 | 1.6582 | 0.84393 | 7.6908 |
| 0.8 | 0.72730 | 0.73856 | 1.5471 | 0.77976 | 7.2113 |
| 0.9 | 0.66832 | 0.67799 | 1.4464 | 0.71362 | 6.7784 |
| 1 | 0.60804 | 0.61628 | 1.3548 | 0.64686 | 6.3846 |

TABLE VI
LOWER BOUND COMPARISON: a=0.1, b≥a

| b | Q1(a,b) | LB1JP | LB1JP (ε %) | LB1A | LB1JP (ε %) |
|---|---|---|---|---|---|
| 0.1 | 0.99503 | 0.99338 | 0.1664 | 0.12408 | 87.5293 |
| 0.2 | 0.98029 | 0.97866 | 0.1664 | 0.22615 | 76.9303 |
| 0.3 | 0.95621 | 0.95462 | 0.1663 | 0.30711 | 67.8826 |
| 0.4 | 0.92348 | 0.92194 | 0.1663 | 0.36822 | 60.1263 |
| 0.5 | 0.88304 | 0.88157 | 0.1663 | 0.41105 | 53.4499 |
| 0.6 | 0.83602 | 0.83463 | 0.1662 | 0.43740 | 47.6802 |
| 0.7 | 0.78366 | 0.78235 | 0.1662 | 0.44923 | 42.6752 |
| 0.8 | 0.72731 | 0.72616 | 0.1661 | 0.44862 | 38.3174 |
| 0.9 | 0.66832 | 0.66721 | 0.1660 | 0.43768 | 34.5097 |
| 1 | 0.60804 | 0.60703 | 0.1660 | 0.41851 | 31.1712 |

TABLE VII
UPPER BOUND COMPARISON: a=2, b<a

| b | Q1(a,b) | UB2JP | UB2JP (ε %) | UB2A | UB2A (ε %) |
|---|---|---|---|---|---|
| 1 | 0.91810 | 0.91883 | 0.0797 | 0.93517 | 1.8587 |
| 1.1 | 0.89807 | 0.89913 | 0.1179 | 0.91929 | 2.3626 |
| 1.2 | 0.87533 | 0.87701 | 0.1914 | 0.90118 | 2.9528 |
| 1.3 | 0.84985 | 0.85254 | 0.3164 | 0.88081 | 3.6426 |
| 1.4 | 0.82164 | 0.82584 | 0.5118 | 0.85819 | 4.4488 |
| 1.5 | 0.79076 | 0.79708 | 0.7989 | 0.83340 | 5.3922 |
| 1.6 | 0.75736 | 0.76647 | 1.2019 | 0.80658 | 6.4981 |
| 1.7 | 0.72164 | 0.73426 | 1.7486 | 0.77791 | 7.7979 |
| 1.8 | 0.68386 | 0.70076 | 2.4715 | 0.74767 | 9.3301 |
| 1.9 | 0.64436 | 0.66632 | 3.4089 | 0.71615 | 11.1419 |
| 2 | 0.60350 | 0.63130 | 4.6076 | 0.68371 | 13.2922 |

TABLE VIII
LOWER BOUND COMPARISON: a=20, b<a

| b | Q1(a,b) | LB2JP | LB2JP (ε %) | LB2A | LB2A (ε %) |
|---|---|---|---|---|---|
| 19.1 | 0.82267 | 0.82007 | 0.3161 | 0.80424 | 2.2393 |
| 19.2 | 0.79546 | 0.79235 | 0.3904 | 0.77341 | 2.7711 |
| 19.3 | 0.76591 | 0.76223 | 0.4809 | 0.73971 | 3.4201 |
| 19.4 | 0.73414 | 0.72980 | 0.5909 | 0.70322 | 4.2116 |
| 19.5 | 0.70032 | 0.69524 | 0.7249 | 0.66406 | 5.1770 |
| 19.6 | 0.66467 | 0.65877 | 0.8879 | 0.62243 | 6.3550 |
| 19.7 | 0.62748 | 0.62066 | 1.0865 | 0.57858 | 7.7936 |
| 19.8 | 0.58906 | 0.58123 | 1.3287 | 0.53279 | 9.5528 |
| 19.9 | 0.54976 | 0.54083 | 1.6246 | 0.48540 | 11.7077 |
| 20 | 0.50997 | 0.49984 | 1.9869 | 0.43677 | 14.3531 |